# Single photon counting imaging system via compressive sensing

Wen-Kai Yu, [1, 2, 3] Xue-Feng Liu, [1, 2, 3] Xu-Ri Yao, [1, 2, 3] Chao Wang, [1, 2]
Shang-Qi Gao, [1, 2, 3] Guang-Jie Zhai, [1, 2, *] Qing Zhao, [4] Mo-Lin Ge, [4]

[1]*Laboratory of Space Science Experiment Technology, Center for Space Science and Applied Research, Chinese Academy of Sciences, Beijing, 100190, China*
[2]*Laboratory of Space Science Experiment Technology, National Space Science Center, Chinese Academy of Sciences, Beijing, 100190, China*
[3]*Graduate University of Chinese Academy of Sciences, Beijing, 100190, China*
[4]*College of Physics, Beijing Institute of Technology, Beijing, 100081, China*
*\*gjzhai@nssc.ac.cn*

**Abstract:** An imaging system based on single photon counting and compressive sensing (ISSPCCS) is developed to reconstruct a sparse image in absolute darkness. The single photon avalanche detector and spatial light modulator (SLM) of aluminum micro-mirrors are employed in the imaging system while the convex optimization is used in the reconstruction algorithm. The image of an object in the very dark light can be reconstructed from an under-sampling data set, but with very high SNR and robustness. Compared with the traditional single-pixel camera used a photomultiplier tube (PMT) as the detector, the ISSPCCS realizes photon counting imaging, and the count of photons not only carries fluctuations of light intensity, but also is more intuitive.

©2012 Optical Society of America

## 1. Introduction

In recent years, single photon detector has drawn much attention and played an important role in many applications [1,2]. Particularly due to its single-photon precision, and discrete photon accounting instead of light intensity, therefore it is indispensible to be used in the ultra dark environment. Attributable to the high sensitivity and excellent time resolution [3,4], photon counting imaging with a single photon detector is widely used in various fields, such as quantum imaging [5-7], fluorescence lifetime microscopy [8] and laser radar [9]. To get an image with spatial resolution, single photon detector array is required. Unfortunately, single photon detector array is still under study so far because of its difficulties in fabrication and limitation in performance [10,11]. Even to the existing single photon detector arrays, the number of pixels is usually too small to provide useful spatial resolution [12]. To solve this problem, a prevalent way is to scan a single photon point to achieve spatial resolving power. However, it wastes a lot of time and is inconvenient to carry out.

Fortunately, a new image reconstruction theory, compressed sensing (CS), was rigorously formulated to systematically and accurately reconstruct a sparse image from an undersampled data set in 2005 [13-18]. That means images can be reconstructed from far fewer measurements than what is usually considered necessary by Shannon/Nyquist sampling theorem (Nyquist 1928; Shannon & Weaver 1949). CS has been applied in many fields, as it dramatically reduces the sampling number in signal detection [19-23]. Based on this theory, a novel imaging system named single-pixel camera is pioneered by Richard G. Baraniuk [24-26]. This technique can reconstruct images with just a single-pixel detector without scanning it. On behalf of its particular features, the single-pixel camera has been used in a number of imaging systems such as compressive color measurements [27,28] and 3D laser radar [29,30]. The conventional prototype of the single-pixel camera employs a digital micromirror device (DMD) to perform linear projections of an image onto pseudorandom binary patterns, and a photomultiplier tube (PMT) is used to collect the light reflecting from the DMD at a certain direction. With the foreknown random measurement matrix on the DMD and information of light intensity detected by PMT, the image of an object can be constructed via the algorithm of compressive sensing.

In this paper, we combine photon counting imaging with compressed sensing theory to achieve photon counting imaging with only one single photon point detector. We still use a DMD to create the modulation of the image, and the photons reflected from the DMD are detected by a Geiger-mode silicon avalanche photodiode, which is well developed right now for single photon counting. Using this system, we achieved imaging of several objects under ultra weak illumination and obtained exciting results.

## 2. Compressive sensing

According to the Nyquist-Shannon sampling theorem, the correct sampling of a band limited signal should be done at a rate at least twice the bandwidth. Only by this way can all the frequencies in the bandwidth be observed. The CS theory has shown that a small collection of nonadaptive linear measurements of a compressible signal or image contain enough information for reconstruction and processing [13-18]. The interesting thing is that, in many instances, natural signals have a structure, which is sparse or sparse under representation of certain basis set. Under this precondition, we can measure a few linear projections of the signal and then reconstruct it by solving a convex optimization problem. One or more measurements can be lost without corrupting the entire reconstruction [24].

We represent observed samples using the column vector $x$ with elements $x[n]$, with $n = 1, 2……N$. Instead of measuring the signal itself, we measure the scalar product of the signal with a series of row vectors [31,32]:

$$y = Ax + e ,\qquad(1)$$

where the matrix $A$ contains $M$ row vectors with dimension $N$, forming a $M \times N$ measurement matrix, which is related to the measurement approach. The vector $y$ is the measurement vector with dimension $M$, and $e$ is the noise vector.

In the traditional measurement, the number of scalar products measured must be equal to the dimension of the signal ($M = N$). Consequently, it is straightforward to recover the vector $x$ providing that rank($A$) = $N$. The basic idea in CS image reconstruction theory is that the vector $x$ can be recovered from $y$ and $A$ even when $M < N$. Since this problem is an ill-conditioned, careful choice of $A$ and prior knowledge about the signal $x$ are required in the recovering process [17].

On the one hand, the target signal must be sparse or sparsified in a certain basis set, such as discrete cosines or Haar wavelets. We assume the basis of $N \times 1$ vectors $\{\psi_i\}_{i=1}^{N}$ is orthogonal, and then the signal $x$ can be represented in terms of the $N \times N$ basis matrix $\Psi = [\psi_1, \psi_2, ..., \psi_N]$:

$$x = \sum_{i=1}^{N} x_i' \psi_i \quad \text{or} \quad x = \Psi x', \tag{2}$$

where $x'$ is the $N \times 1$ column vector of the projection of $x$ on $\Psi$. The signal $x$ is $K$ sparse if there is only $K$ of the projection $x_i'$ in Eq. (2) is nonzero. Compressive sensing proves that the signal is compressible if it is $K$ sparse in some certain basis. Our problem becomes a problem of solving the equation:

$$y = A\Psi x' + e. \tag{3}$$

If we can determine the positions and values of the $K$ nonzero $x'$, then the original signal $x$ can be recovered from Eq. (2).

On the other hand, the measurement matrix $A$ must be incoherent with the sparse basis $\Psi$, which means the sparse basis $\Psi$ should not be represented by measurement matrix $A$ [14-16]. An important conclusion is that random matrices are largely incoherent with nay fixed basis $\Psi$ [33]. This dramatically improves the broad application of CS theory.

As rank($A\Psi$) $<< N$, the number of equations is smaller than the number of unknown. An appropriate solution to Eq. (3) is to look for the vector with famous $l_2 - l_1$ problem (called basis pursuit denoising (BPDN)) [34,35]:

$$\min_{x'} \frac{1}{2}\|y - A\Psi x'\|_2^2 + \tau \|x'\|_1, \tag{4}$$

where $\tau$ is a constant scalar, and $\|\cdots\|_p$ stands for the $l_p$ norm, defined as $(\|x\|_p)^p = \sum_{i=1}^{N} |x_i|^p$. It is proven that a $K$-sparse vector can be exactly recovered using just $M \leq O(K \cdot \log(N/K))$ random measurements. When the signal has large sparseness, the required measurements should be much fewer than traditional required.

Here, we use Sparse Reconstruction by Separable Approximation (SpaRSA) to reconstruct the signal. SpaRSA is an algorithmic framework for problems like Eq. (1), which can be instantiated by adopting different rules, different ways to choose parameters, and different criteria to accept a solution to each subproblem [36].

## 3. Single photon counting imaging system via compressive sensing

The schematic diagram of the experimental apparatus is given in Fig. 1. Light from a halogen lamp with 55W power passes the object and then is attenuated by several neutral density filters into the single photon level. This simulates an object under ultra weak illumination. The object is then imaged on the DMD by an imaging lens. The DMD array, based on Micro Electro and Mechanical System (MEMS), is a product consisting of a 1024 × 768 micro-mirrors. The size of a micro-mirror is only 13.68μm × 13.68μm, and the total size of the active mirror array is about 14mm × 10.5mm. Each mirror rotates about a hinge and can be positioned in one of two states, +12 degrees and -12 degrees away from initial position. Thus light falling on the DMD may be reflected in two directions depending on the orientation of the mirrors, which is decided by a matrix loaded into DMD [12]. The light reflecting from the mirrors stated +12 degrees is focused on a Geiger-mode silicon avalanche photodiode. A fiber collimator and several lenses are used to increase the couple efficiency. The avalanche photodiode counts the number of photons collected, which can be thought to be approximately proportional to the intensity of light reflected in this direction.

In the imaging system, the DMD performs a linear projection on the image. The matrix loaded from the DMD corresponds to the measurement matrix of $A_i$. The measurement result $y_i$ is obtained by the photon counts on the single photon detector. In order to image an object, *M* random binary patterns generated by a random number generator on the control platform are sequentially loaded on the DMD, driving the micro mirrors changing their positions at the frequency of 34Hz. The single photon detector then records *M* photon counts correspondingly. The row of matrixes *A* from DMD and corresponding results *y* from the single photon detector are then sent to the algorithm unit. To solve the problem of Eq. (4), we use SpaRSA algorithm for image reconstruction.

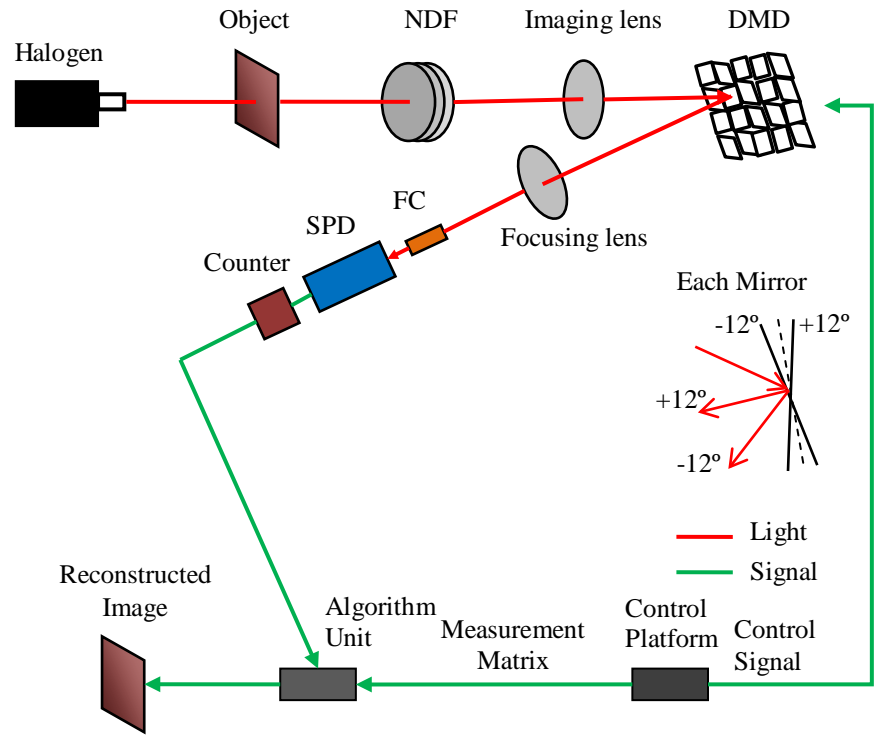

Fig. 1. Experimental apparatus for single photon counting imaging system via compressive sensing. Halogen: halogen lamp, NDF: neutral density filters, DMD: digital micromirror device, FC: fiber collimator, SPD: single photon detector, Counter: counter board. The light path is shown with red lines, while electronic signal with green lines. The inset shows the detailed structure of DMD.

Fig. 2 presents the experimental setup comprising the optical hardware for the photon counting compressive sensing imaging system. Light path has been expressed with red lines. We can see that the system is compact, occupying the area less than 0.2 square meters. With fine design, the system is easy to be miniature and forms steady equipments.

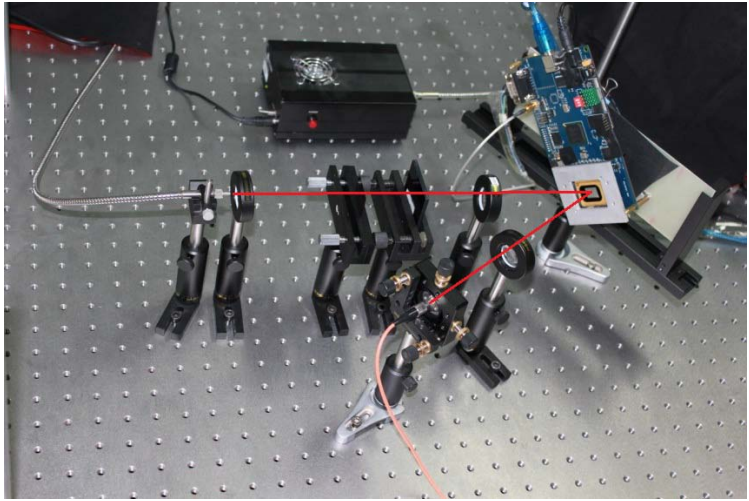

Fig. 2. Optical setup of single photon counting imaging system via compressive sensing. Light path has been expressed with red lines.

## 4. Experimental results and discussion

*4.1. NSSC*

Several objects were imaged to validate the system's capabilities. In the first experiment, the image of a film printed with NSSC (the abbreviation of National Space Science Center) is reconstructed. The size of each letter is about 1mm × 1mm, which is imaged in 64 × 64 regions on the DMD. The four letters were imaged one by one. First, the micro-mirrors were set in the 64 × 64 regions, the letter "N" work area, while the light on the other micro-mirrors will not be reflected into the detector, which can be accomplished by the control platform of the image system. Following the measurement method mentioned above, 1800 random 64 × 64 patterns were used to image the first letter "N". After 53s for 1800 measurements, the random pattern $A$ and photon count $y$ were sent into SpaRSA algorithm unit, which then produced the 64 × 64 pixel image of letter "N". In the algorithm unit, the wavelets were used as sparse basis, and SpaRSA algorithm for image reconstruction. Then the working region of the DMD was changed, and the other letters were imaged. Finally, the images of the four letters were spliced together to form a 64 × 256 pixel image of the 'NSSC' film presented in Fig. 3. The significant point is that the number of measurement (1800) is just smaller than half of the total pixel image (4096), which is impossible to success in traditional imaging. In fact, even when the number of measurement $M$ is reduced to 600, the letters can still be reconstructed, but in lower resolution.

In the process of reconstruction, the average number of photons recorded by the single photon detector is about $5.61 \times 10^5$ per second. The counting rates of all the measurements in imaging the four letters are averaged, and then the four counting rates are added together. That means the total number of photons emitted from 4096 pixels is $5.61 \times 10^5$ per second, and therefore, about 34 photons per second from one pixel. The frequency used on the DMD is 34Hz, which means there is only one photon emitted from each pixel of the image in each random pattern, or in a single measurement. The amazing result reveals that the light intensity from the object is in the single photon level, and our imaging system has achieved the ability to work in ultra weak illumination.

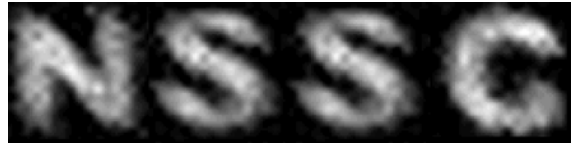

(a)

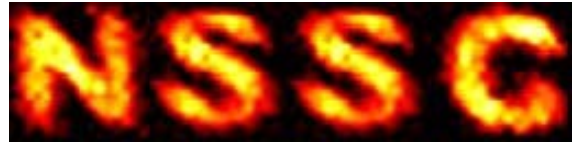

(b)

Fig. 3. (a) The gray image of abbreviation of National Space Science Center (NSSC). (b) The pseudo-color map of abbreviation of National Space Science Center (NSSC). Size of the image is 64 × 256 pixels. The 1800 measurements are used to get the image. The photon energy collected by this image is about 0.02nJ, and the average number of photons is $5.61 \times 10^5$ per second.

*4.2. BIT*

Despite the progress of various CS algorithms in accurate reconstruction, the size of reconstructed image is still limited. Because of the enormous memory occupation and long running time, CS is restricted in practical applications. To reconstruct a large pixel image, a method of DMD work area scanning was advanced. After coupling photons from the whole area of the DMD well into the single photon detector, the working area of DMD in the first row was set, and scanned row by row. Then the measurement steps mentioned above was conducted and repeated. The scanning of working area can be carried out easily by the control platform while the entire optical hardware needn't to be adjusted in the scanning process. The image of each row was reconstructed by SpaRSA algorithm separately. Finally, the images of rows were spliced sequentially together to form an image of 1024 × 768 pixels. An image on the full screen of DMD can be reconstructed by this method. Because the CS algorithm reconstructs the image of one row each time, the running time is acceptable. It must be emphasized that the scanning of DMD work area is totally different from traditional detector scanning. First of all, the scanning of DMD work area is achieved by software without any mechanical movement; therefore, it is highly accurate and stabile. Secondly, the image of one row can be obtained in each scanning instead of a pixel. Consequently, it is obviously more efficient than the scanning of traditional detector.

The image of a film printed with the abbreviation of Beijing Institute of Technology (BIT) was reconstructed as shown in Fig. 4. We scanned 79 rows to cover the three letters. 600 random patterns were used for each row of the DMD, and the size of the image is 1024 × 79 pixels. Clearly, this image has higher spatial resolution and sharper edge because of the more pixels. The result suggests that the image system work very well with measurement of DMD work area scanning. It also reveals that even $10^4$ photons per second are enough for image reconstruction, which is much less than that used in the reconstruction of the letter "NSSC". The experiment to quantify the number of photons required was carried out with the reduction of the photon counting via attenuating the light more intensively. A very interesting result obtained is that although the number of photons is decreasing, the reconstructed image with the same number of measurement is still clearly observable.

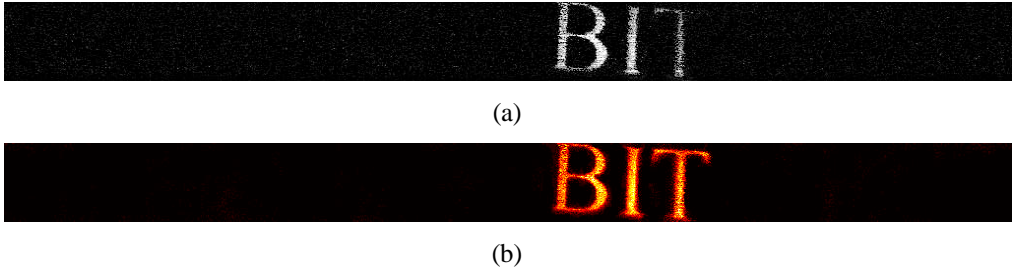

(a)

(b)

Fig. 4. (a)The gray image of abbreviation of Beijing Institute of Technology (BIT). (b) The pseudo-color map of abbreviation of Beijing Institute of Technology (BIT) with image processing. Size of the image is 1024 × 79 pixels. The total count of photons used to reconstruct the image of object is $9.2139 \times 10^7$. Thus the photon energy collected by this image is about 0.0387nJ, and the average number of photons is $6.6312 \times 10^4$ per second.

*4.3. SNR*

In order to further assess the signal to noise ratio (SNR) of the system, two pieces of data measured in the same sequence of random patterns placed on the DMD are plotted for comparison in Fig. 5. *Data1* in blue line and *data2* in red line are the photon counts in the same condition. Without any noise, these two sequences of photon counts should match exactly. However, because of the noise existing in the system, there are slight errors between the two sequences. The deviation between *data1* and *data2*, which has been already added the mean value of *data1*, is shown in green line in Fig. 5. This deviation can be treated as the system noise. In our experiment, if the fluctuation of data is overrun by the fluctuation of noise, the image cannot be reconstructed. Thus, the fluctuation of the *signal* and *noise* is more important than the absolute value. The SNR of the system can be defined as

$$SNR = 10 \lg \frac{signal}{noise}$$

$$= 10 \lg \frac{\mathrm{var}(data1) + \mathrm{var}(data2)}{\mathrm{var}(deviation)}, \qquad (5)$$

where the var is the variances of *date1*, *date2* and the difference, which are $2.92 \times 10^5$, $2.80 \times 10^5$, and $3.46 \times 10^4$, respectively. The SNR in the experiment is up to 12.2 dB.

In fact, there is still space for photons used to decrease, until the fluctuation of signal is near that of noise. The noise of the system contains both the intrinsic dark counts of single photon detector and the light noise from the environment. With the well designed optical system, the noise from the environment can be nearly eliminated. Therefore, the fluctuation of dark counts of the single photon detector (about several hundred per second) is the baseline of the weakness of object illumination.

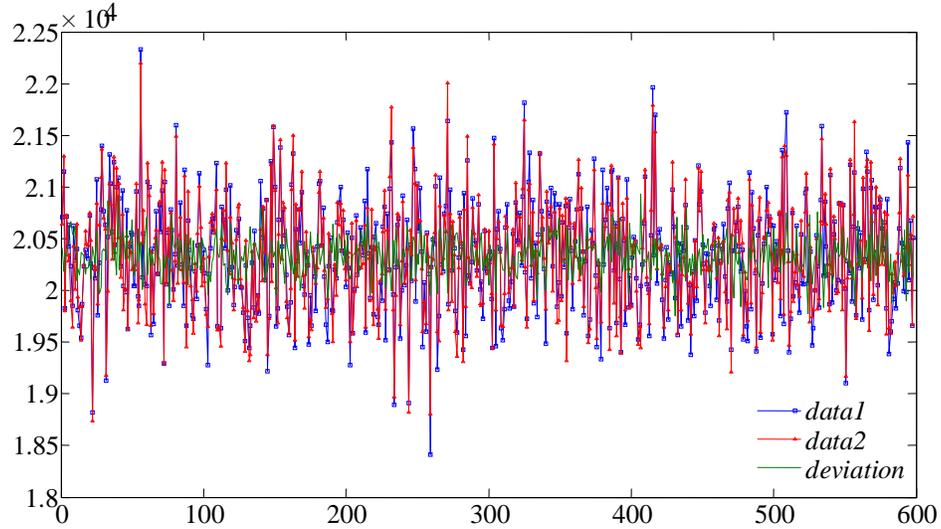

Fig. 5. The SNR analysis of the system. *Data1* in blue line and *data2* in red line are the photon counts in the condition of same sequence of random patterns. The *deviation* between *data*1 and *data2*, which has been already added the mean value of *data1*, is shown in green line. The point on horizontal axis corresponds to how many times it measures, and the vertical axis is the count of photons in each measurement.

## 5. Conclusion

A novel single photon counting imaging system based on single photon detector and compressive sensing is presented in this paper. The experiments demonstrated that the imaging system can reconstruct the object image under the illumination of single photon level with under sampling and good quality. The more interesting results revealed that the reconstructed image with the same number of measurements is still clearly visible in spite of the decreasing of the photon number. As a consequence of the high sensitivity, low sampling and high SNR of the imaging system, it has a great prospect in biology fluorescence, material structure and other related fields.

## Acknowledgments

This work was supported by the National High Technology Research and Development Program of China under Grant Project No. 2011AA120102.